\documentclass[twocolumn,showpacs,preprintnumbers,amsmath,amssymb]{revtex4}

\usepackage{epsfig,graphicx,dcolumn,bm,times}

\begin{document}

\title{Rank-dependent deactivation in network evolution}

\author{Xin-Jian Xu$^{1,2,}$\footnote{Electronic address: xinjxu@shu.edu.cn} and Ming-Chen Zhou$^{1,}$\footnote{Electronic address: harrymczhou@hotmail.com}}

\address{$^{1}$Department of Mathematics, College of Science, Shanghai University, Shanghai 200444, China\\$^{2}$Institute of Systems Science, Shanghai University, Shanghai 200444, China}

\date{\today}

\begin{abstract}
A rank-dependent deactivation mechanism is introduced to network
evolution. The growth dynamics of the network is based on a finite
memory of individuals, which is implemented by deactivating one site
at each time step. The model shows striking features of a wide range
of real-world networks: power-law degree distribution, high
clustering coefficient, and disassortative degree correlation.
\end{abstract}


\maketitle

Since 1998, with the small-world model introduced by Watts and
Strogatz \cite{WS98}, we have witnessed the emergence of a new
science of networks, which has powerful function of describing
structures \cite{AB02,CRTB07} and dynamics \cite{BLMCH06,DGM08} of
many real systems. A network is a mathematical object which consists
of vertices connected by edges. Despite differences in their nature,
most real-world networks are characterized by similar topological
properties, in contrast to those obtained by traditional random
graphs. For instance, real networks display higher clustering than
that expected from random networks \cite{ASBS00}. Also, it has been
found that many large networks are scale free (SF) \cite{BA99},
which means a power-law distribution of connectivity, $P(k)\sim
k^{-\gamma}$, where $P(k)$ is the probability that a vertex in the
network is of degree $k$ and $\gamma$ is a positive real number
determined by the given network. In order to understand how SF
networks arise, Barab\'{a}si and Albert (BA) proposed an evolving
network model in 1999, which grows at a constant rate and new
vertices attach to old ones with probability $\Pi(k) \sim k$
\cite{BA99}. In this way, vertices of high degree are more likely to
receive further edges from newcomers.

For most networks, however, aging of sites usually occurs. For
instance, in reference networks old papers are rarely cited; in
social networks people of the same age are more likely to be
friends. To study the effect of aging on network evolution, the BA
model has been modified by incorporating time dependence in the
network \cite{DM00,ZWZ03,HS04,HAB07,RL07,DKM08,CM09}. Dorogovtsev
and Mendes studied the case when the connection probability of the
new site with an old one is not only proportional to the degree $k$
but also to the power of its present age $\tau^{-\alpha}$
\cite{DM00}. They showed both numerically and analytically that the
scale-free nature disappears when $\alpha < -1$. As an alternative,
Zhu et al introduced the exponential decay function $e^{-\beta\tau}$
of its present age to the BA attachment probability \cite{ZWZ03}. It
was found that the produced network is significantly transformed
besides the change in the degree distribution. On the other hand,
Klemm and Egu\'{i}luz \cite{KE02a} observed the negative correlation
between the vertex age and its rate of acquiring links from the
network of scientific citations. Based on that, they investigated
the finite collective memory of popular individuals and proposed a
highly clustered scale-free network model \cite{KE02a,KE02b}. The
model accounts for three empirical features: preferential
attachment, power-law degree distribution, and negative correlation
between age and connection rate.

Recently, Fortunato et al proposed a criterion of network growth
that explicitly relies on the ranking of vertices \cite{FFM06},
which originates from the idea that the absolute importance
(popularity or fitness) of an object is often difficult or
impossible for strangers to measure in social networks. Instead, it
is quite common to have a clear knowledge about the relative values
of two objects, i.e., who is more popular or richer between two
individuals. The rank-driven mechanism generates networks with the
scale-free degree distribution when the probability to link a target
vertex is any power-law function of its rank, even when one has only
partial information of vertex ranks \cite{FFM06,TS07}. Since the
perception of how items are ranked requires far less information
than their actual importance, the rank-driven mechanism can well
mimic the reality in many cases that the relative values of agents
are easier to access than their absolute values. In this paper, we
integrate rank with deactivation and study their influences on
network evolution. Simulations show that interesting statistical
properties of the generated network display good features observed
in realistic systems.

The present model is based on the rank-dependent deactivating of
vertices, which describes the growth dynamics of a network with
directed links, run as follows. First, start from an initial network
of $m$ completely connected seeds, whose states are active. By
$k^{\text{in}}$ we denote the in-degree of the vertex, i.e., the
number of edges pointing to the vertex. At each time step, add a new
vertex $n$ with $m$ outgoing edges. The new vertex is disconnected
at first, so $k_{n}^{\text{in}}=0$ at this point. Each vertex $i$ of
the $m$ active vertices receives exactly one incoming edge, thereby
$k_{i}^{\text{in}} \rightarrow k_{i}^{\text{in}} +1$. Then activate
the new vertex $n$ and deactivate one (denoted by $j$) of the $m+1$
active vertices with probability
\begin{equation}
\Pi(j)=\frac{\gamma}{a+R_j}, \label{deactprob}
\end{equation}
where $a \ge 0$ is a constant bias and $\gamma$ is the normalization
factor. $R_j \in [1, m+1]$ is the rank of $j$ among the $m+1$ active
vertices. The average connectivity of the network is given by the
number $m$ of outgoing edges per vertex. The new added vertex is
always in the active state first and receives edges from
subsequently added vertices until it is deactivated. Note that the
larger rank a vertex possesses, the more difficult for it to be
deactivated. For the case of the citation network, Eq.
(\ref{deactprob}) means that the famous paper cited mostly is less
possibility to be forgotten. In Ref. \cite{FFM06}, the model grows
according to the rank-based preferential attachment
$\Pi(n\rightarrow i) \sim R_{i}^{-\alpha}$. In case vertices are
sorted by age $R(t)=t$, the older the vertex is, the higher
possibility for it gaining new edges, coinciding with ours.

\begin{figure}
\includegraphics[width=\columnwidth]{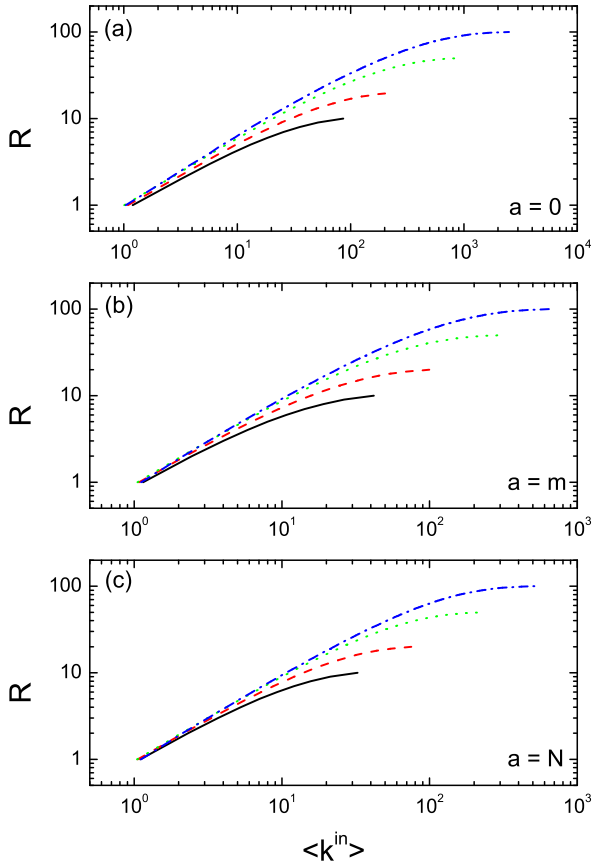}

\caption{(Color online) Illustration of average $R$ versus average
$k^{\text{in}}$ of the generated networks for various values of $m$:
$m=10$ (solid line), $20$ (dashed line), $50$ (dotted line), and
$100$ (dashed dotted line), respectively. The size of the network is
$N=10^{5}$.} \label{fig1}
\end{figure}

The choice of prestige measure can be arbitrary, either topological
measures or physical ones. In present work we sort vertices by age
for simplicity. Namely, the older the vertex is, the larger rank it
possesses. Supposing the distribution of the in-degree of active
vertices at time $t$ denoted by $p(k^{\text{in}},t)$, then we can
write out the master equation
\begin{equation}
p(k^{\text{in}}+1,t+1)=[1-\Pi(k^{\text{in}})]p(k^{\text{in}},t),
\label{masterk}
\end{equation}
where $\Pi(k^{\text{in}})$ is the deactivation probability of a
vertex with in-degree $k^{\text{in}}$. To do further calculation, we
should get the relation between $R$ and $k^{\text{in}}$. In Fig.
\ref{fig1} we plot $R$ as a function of $k^{\text{in}}$ of the model
by numerical simulations. In order to reduce statistical error, the
in-degrees of the vertices are calculated as an average. As it can
be seen, there is a rough power law between $R$ and $k^{\text{in}}$,
$R\sim\mu(k^{\text{in}})^{\nu}$ \cite {Note1}. Then one easily
obtains $\Pi(k^{\text{in}})\sim\mu\nu(k^{\text{in}})^{\nu-1}\Pi(R)$,
where $\Pi(R)$ is the deactivation probability of a vertex with rank
$R$. Substituting them into Eq. (\ref{masterk}), we get
\begin{eqnarray}
p(k^{\text{in}}+1,t+1)
&=&[1-\mu\nu(k^{\text{in}})^{\nu-1}\Pi(R)]p(k^{\text{in}},t) \notag\\
&=&\left[1-\frac{\gamma\mu\nu(k^{\text{in}})^{\nu-1}}{a+\mu(k^{\text{in}})^{\nu}}\right]p(k^{\text{in}},t).
\label{masterr}
\end{eqnarray}

\begin{figure}
\includegraphics[width=\columnwidth]{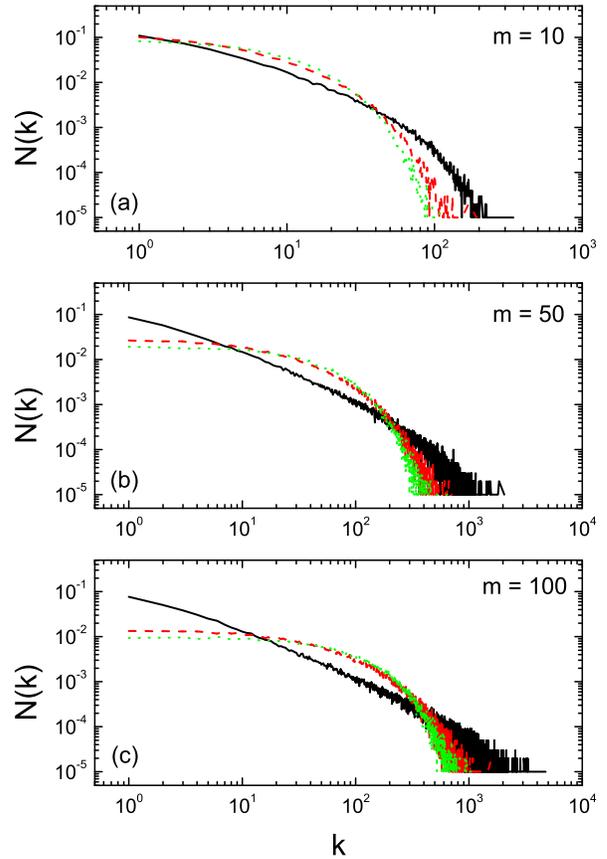}

\caption{(Color online) Log-log plots of the vertex degree
distribution of the generated networks for different values of $a$:
$a=0$ (solid line), $m$ (dashed line), and $N$ (dotted line),
respectively. All the experiment networks have a total number of
vertices $N=10^{5}$.} \label{fig2}
\end{figure}

The subsequent thing is just to follow the analytical method in Ref.
\cite{KE02a}. Imposing the stationarity condition
$p(k^{\text{in}},t+1)=p(k^{\text{in}},t)$ yields
\begin{equation}
p(k^{\text{in}}+1)-p(k^{\text{in}})=-\frac{\gamma\mu\nu(k^{\text{in}})^{\nu-1}}{a+\mu(k^{\text{in}})^{\nu}}p(k^{\text{in}}).
\label{recursive}
\end{equation}
Assuming $k^{\text{in}}$ changes continuously, we have
\begin{equation}
\frac{dp(k^{\text{in}})}{dk^{\text{in}}}=-\frac{\gamma\mu\nu(k^{\text{in}})^{\nu-1}}{a+\mu(k^{\text{in}})^{\nu}}p(k^{\text{in}}),
\label{differpk}
\end{equation}
and accordingly obtain the solution
\begin{equation}
p(k^{\text{in}})=b[a+\mu(k^{\text{in}})^{\nu}]^{-\gamma},
\label{solupk}
\end{equation}
with the appropriate normalization constant $b$. In case the total
number $N$ of vertices in the network is large compared with the
number $m$ of active vertices, the overall in-degree distribution
$N(k^{\text{in}})$ can be calculated as the rate of the change of
the in-degree distribution $p(k^{\text{in}})$ of the active
vertices, which obeys
\begin{equation}
N(k^{\text{in}})=-\frac{dp(k^{\text{in}})}{dk^{\text{in}}}=b\gamma\mu\nu(k^{\text{in}})^{\nu-1}[a+\mu(k^{\text{in}})^{\nu}]^{-\gamma-1}.
\label{solunk}
\end{equation}
If we chose the value of the bias $a=0$, Eq. (\ref{solunk}) is no
other than the probability distribution of the total degree
$k=k^{\text{in}}+m$ of vertices
\begin{equation}
N(k)=b\gamma\mu^{-\gamma}\nu k^{-\nu\gamma-1}.
\end{equation}
In Fig. \ref{fig2} we plot the total degree distribution of the
resulting networks with different values of $a$. All the plots
display the good right-skewed behavior, which is reasonably in
agreement with the condition of many realistic systems
\cite{WM00,MN01}. Especially for $a=0$, we notice beautiful power
laws.

\begin{figure}
\includegraphics[width=\columnwidth]{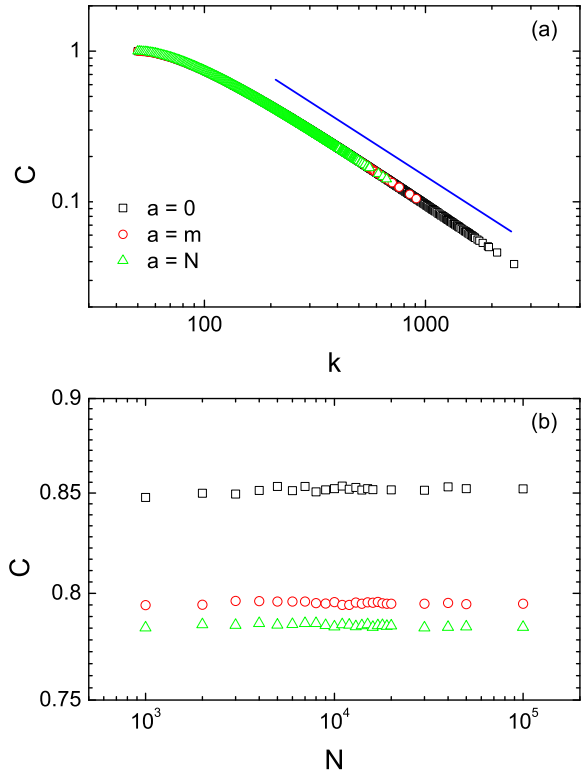}

\caption{(Color online) Average clustering coefficient $C$ as a
function of $k$ (a) and $N$ (b), respectively. The parameter $m=50$
for all plots.} \label{fig3}
\end{figure}

The \lq\lq complexity\rq\rq of networks usually cannot be fully
characterized by the degree distribution of vertices. Instead, the
self-organization of structures of complex networks is
mathematically encoded in various correlations existing among
different vertices. To describe the network structure in more
detail, several other topological quantities have been introduced to
the statistics of networks, such as clustering coefficient, degree
correlation, shortest path length, and so on. In the following, we
shall go beyond the degree distribution and discuss those
quantities.

Let us start with the clustering coefficient $C$ of the network,
which is defined as the average probability with which two neighbors
of a vertex are also neighbors to each other. For example, if a
vertex $i$ has $k_i$ edges, and among its $k_i$ nearest neighbors
there are $e_i$ edges, then the clustering coefficient of $i$ is
defined by
\begin{equation}
C_i=\frac{2e_i}{k_i(k_i-1)}. \label{defcluster}
\end{equation}
In order to compute the clustering coefficient, we shall consider
the network as undirected and denote by $k=k^{\text{in}}+m$ the
total degree of vertex $i$. In the deactivation model, new edges are
created between the active vertices and the added one. Moreover, all
the active vertices are connected. At each time step, the degree
$k_i$ of each active vertex $i$ increases by $1$ and $e_i$ increase
by $m-1$. Therefore, the evolutionary dynamics of $k_i$ and $e_i$
are given by
\begin{eqnarray}
k_i&=&m+t, \label{evolutionk}\\
\frac{de_i}{dt}&=&m-1. \label{evolutione}
\end{eqnarray}
Integrating Eq. (\ref{evolutione}) with the boundary condition
$e_i(0)=m(m-1)/2$ and substituting the solution into Eq.
(\ref{defcluster}), we recover the clustering coefficient $C(k)$
restricted to the vertices of degree $k$ \cite{KE02b,VBMPV03},
\begin{equation}
C(k)=\frac{2(m-1)}{k-1}-\frac{m(m-1)}{k(k-1)}. \label{expressc}
\end{equation}
The expression indicates that the local clustering scales as $C(k)
\sim k^{-1}$ for large $k$. In Fig. \ref{fig3}(a) we plot the
average clustering coefficient $C(k)$ as a function of the vertex
degree $k$. The best linear fit gives $C(k) \sim k^{-\xi}$ with
exponent $\xi=0.98(6)$, which coincides with the prediction of Eq.
(\ref{expressc}). This is the signature of a nontrivial architecture
in which low-degree vertices generically belong to well
interconnected communities while high-degree ones are linked to many
sites that may belong to different groups which are sparsely
connected. In Fig. \ref{fig3}(b) we present three typical curves of
the average clustering coefficient $C$ versus the network size $N$.
It is worth noting that the clustering coefficient of the generated
networks for all cases is higher than that for the corresponding
one-dimensional regular lattices whose value is $3/4$ in the limit
case \cite {Note2}.

\begin{figure}
\includegraphics[width=\columnwidth]{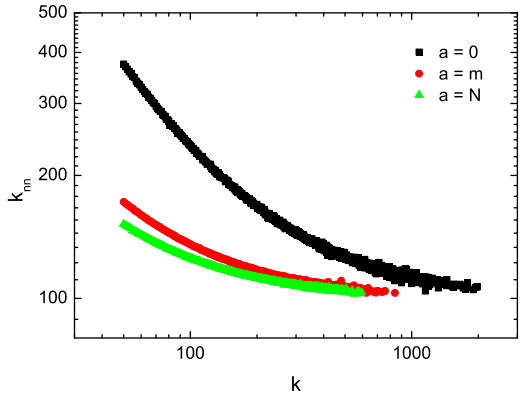}

\caption{(Color online) Average nearest-neighbor degree $k_{nn}$ as
a function of the vertex degree $k$ for different values of the bias
$a$. The network parameters are $N=10^5$ and $m=50$.} \label{fig4}
\end{figure}

Another commonly studied topological quantity is the degree
correlation (or the mixing pattern), which can be characterized by
analyzing the average degree of nearest neighbors, defined by
\cite{PVV01}
\begin{equation}
k_{nn,i}=\frac{1}{k_i}\sum_{j}a_{ij}k_j. \label{defknn}
\end{equation}
If $k_{nn,i}$ does not show any dependence on the degree of $i$, the
network is uncorrelated. In case $k_{nn,i}$ is dependent on the
vertex degree $i$, there are two kinds of correlations. If
$k_{nn,i}$ increases with $k$, the network is assortative mixing;
i.e., vertices with high connectivity will connect preferably to
highly connected ones. If $k_{nn,i}$ decreases with $k$, the network
is disassortative mixing; i.e., vertices with high connectivity will
connect preferably to lowly connected ones \cite{MN02}. In Fig.
\ref{fig4} we show the simulation results of the average degree of
nearest neighbors $k_{nn}$ as a function of $k$ for different values
of $a$. In all cases, the degree correlation in the deactivation
model is disassortative.

\begin{figure}
\includegraphics[width=\columnwidth]{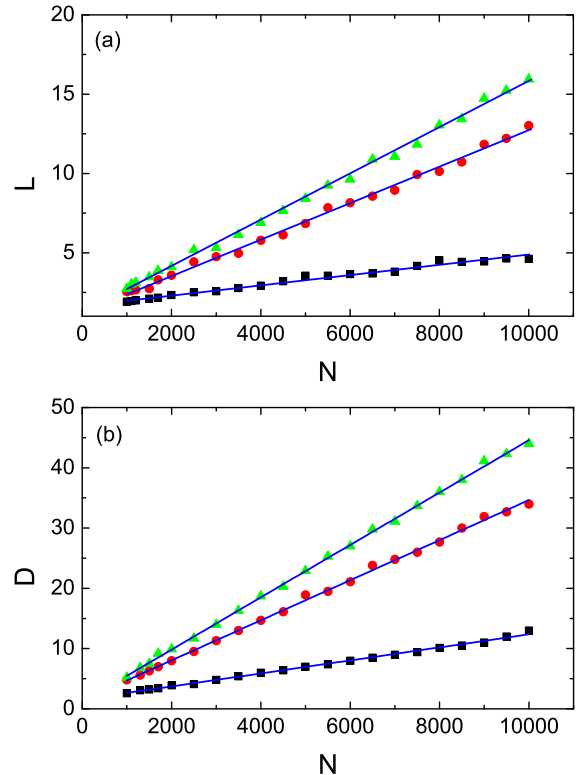}

\caption{(Color online) Scaling of the average shortest path length
(a) and the diameter (b) for different values of the bias $a$: $a=0$
(squares), $m$ (closed circles), and $N$ (triangles), respectively.
The parameter $m=50$ for all plots.} \label{fig5}
\end{figure}

To complete our study of the model, we finally investigate the
scaling of the shortest path length and the diameter of the network.
The shortest path length between two vertices is defined as the
minimum number of intermediate vertices that must be traversed to go
from vertex to vertex. The average shortest path length is the
shortest path length averaged over all the possible pairs of
vertices in the network. On the contrary, the diameter is defined as
the largest among the shortest paths between any two vertices in the
network \cite{VBMPV03}. In Fig. \ref{fig5} we show the scaling
behavior of the average shortest path length $L$ and the diameter
$D$ of the resulting network. Both quantities grow linearly with the
network size $N$ similar to one-dimensional regular lattices. That
is to say, the deactivation model does not exhibit small-world
properties. It indicates that there are very few effective shortcuts
to reduce the distance although the network is highly clustered.
However, the chainlike structure of the model does not necessarily
mean large values of $L$ and $D$. Here $L$ and $D$ are still
relatively small compared with $N$, especially for the case $a=0$.

In summary, we have suggested a rank-dependent deactivation
mechanism and studied its influence on network growth. The resulting
network shows several good features. (i) The degree distribution
$N(k)$ of vertices is power law, which indicates the heterogeneous
topology. (ii) The clustering coefficient is larger than that of
one-dimensional regular lattices. Besides, the local clustering
scales as $C(k) \sim k^{-1}$ for large $k$. (iii) The decay of the
average degree of nearest neighbors $k_{nn}$ with $k$ characterizes
the disassortative mixing pattern. (iv) The average shortest path
length and the diameter grow with the network size, which results
from deactivating sites during network evolution. Most of above
properties have been found very common in realistic systems. We hope
that the measurement conducted in present work could be applied to
real networks in the empirical study. The model we have explored,
however, is possibly the simplest one in the class of rank-dependent
deactivation growing networks. One can notice that in our model the
out-degree of the vertices remain unchanged during the whole
evolving period, which is likely unreasonable for citation networks.
Furthermore, many networks are intrinsically weighted, their edges
having different strengths \cite{BBPV04}. Naturally one can
generalize the present model to the weighted case and sort ranks by
the vertex strength or fitness. There exists a series of
improvements to be made in the future.

The authors acknowledge financial support from NSFC (Grant No.
10805033) and STCSM (Grant No. 08ZR1408000). This work was sponsored
by the Innovation Foundation of Shanghai University.

\end{document}